\documentclass{caosp}
\usepackage{graphicx}
\articleNo{000}
\pubyear{2007}
\volume{37}
\volnumber{2}
\firstpage{1}
\received{Dec 1, 2007}
\accepted{Dec 1, 2007}

\begin{document}

\hauthor{R. Arlt}
\title{The generation and stability of magnetic fields in CP stars}
\author{Rainer Arlt}
\institute{Astrophysikalisches Institut Potsdam, An der Sternwarte 16, D-14482 Potsdam, Germany}
\date{March 8, 2003}

\maketitle

\begin{abstract}
A variety of magnetohydrodynamic mechanisms that may play a role in magnetic, 
chemically peculiar (mCP) stars is reviewed. These involve dynamo mechanisms
in laminar flows as well as turbulent environments, and magnetic instabilities 
of poloidal and toroidal fields as well as combinations of the two. 
While the proto-stellar phase makes the survival of primordial fields
difficult, the variety of magnetic field configurations on mCP stars may
be an indication for that they are instability remnants, but there is
no process which is clearly superior in explaining the strong fields.
\end{abstract}

\section{What do we have to explain?}
A considerable fraction of chemically peculiar A and B
stars (CP stars) show strong surface magnetic fields of about 500~G
up to over 10~kG. The evolution of these fields from 
star formation to their current presence is unknown. This
paper compiles a number of physical processes which may
or may not play a role in the whole scenario of CP star 
magnetism. 

The fraction of main-sequence stars with radiative envelopes 
which show magnetic fields was found to increase from F stars
to late B stars (Wolff 1968). The result was recently confirmed
by Power et al. (2007) who found an increase of mCP star fractions
with stellar mass up to $3.6M_\odot$ for nearby stars. A further 
increase towards higher masses was hidden in the small number of 
higher-mass stars. While the average fraction of mCP stars among
normal stars of the same mass range is generally quoted as being
roughly 10\%, the fraction is less than 2\% in the solar neighbourhood
of 100~pc radius (Power et al. 2007). Why only a fraction of intermediate-mass
stars shows strong magnetic fields is one of the key issues in the search for
their origin. The other is that mCP stars form a distinct sub-group
with slow rotation among A and B stars.

\begin{figure}[t]
\centerline{\includegraphics[width=9cm,clip=]{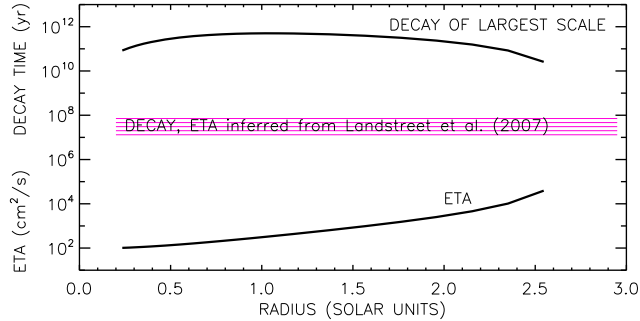}}
\caption{Magnetic diffusivities (lower part of diagram) and the
corresponding decay times for magnetic structures (upper part of 
diagram). The diffusivity depends on the location in the star at 
radius $r$ and so does the decay time of a structure as large as
$r$. The ``small-scale'' curve refers to structures with a scale of $0.1r$
which decay 100~times faster.
A rough estimate of the magnetic diffusivity according to the age
dependence of magnetic field strengths of CP stars is given in
middle.}
\label{decay_times}
\end{figure}

There are indications of a decrease of magnetic field strength with stellar 
age for $M > 3M_\odot$ (Kochukhov \& Bagnulo 2006; Landstreet et al. 2007).
Although the increase of stellar radius may cause the reduction
at constant magnetic flux, it appears that even the flux is decreasing
during the main-sequence life of mCP stars. At first glance this decay 
may be attributed to ohmic decay in the radiative envelope of the star. 
The radial profile of the magnetic diffusivity $\eta(r)$ throughout the 
star according to Spitzer (1956) suggests that the decay time is of the 
order of $10^{11}$~yr everywhere in the radiation zone. The decay time 
inferred from Landstreet et al. (2007) is about $10^7$ to $10^8$~yr (as 
is $\eta$ in cm$^2$/s accidentally). Figure~\ref{decay_times} shows $\eta$ 
as a function of radius and the corresponding ohmic diffusion time 
$\tau_{\rm diff}=r^2/\eta$. The above decay time deduced from observations is also 
indicated -- without radial dependence for obvious reasons. It appears as if the 
magnetic diffusivity is higher than the purely microscopic value hinting 
on very mild turbulence even without convection. 

Another puzzling fact is the decrease of obliquity of the magnetic field  axis 
against the rotation axis with rotation period (Landstreet \& Mathys 2000; 
Bagnulo et al. 2002). MHD processes tend to prefer axisymmetric solutions for 
fast rotators, if a small degree of differential rotation is present in the star. 
This is the very opposite of what is observed. When comparing the rotational 
time-scale with the diffusive time-scale, even the slowest CP stars have 
rotation periods many orders of magnitude shorter than $\tau_{\rm diff}$ and 
are fast rotators in that sense. Since in view of the time-scales normal A stars 
and Ap stars are not vastly different, we tend to place the magnetic fields
first -- the reduced rotation is a consequence of their presence.
In other words, the reason for mCP stars to be magnetic is not their initially
slow rotation.

There is not only the gradual decrease of magnetic field strengths with
the time which the $M>3M_\odot$ stars spent on the main sequence, but also
a possible late emergence of surface fields for stars with $M<2M_\odot$.
Hubrig et al. (2000) found that these stars show magnetic fields only
after they have spent about 30\% of their life on the main sequence. Meanwhile,
counter-examples were found, and the appearance of fields is certainly
not as sharp as it seemed at an earlier research stage.

It should be noted that in general all these characteristics are based on quite 
scattered individual stars. The scatter is very likely not a result
of observing uncertainties. Any physical process leading to the observed 
features must also allow for a certain bandwidth of results or sufficient
input of stochastic behaviour.


\section{The laminar dynamo}
In a largely radiative environment, non-turbulent (laminar) flows 
may still be present, namely meridional circulation and differential 
rotation. These flows are driven by radiatively (Eddington, 1929;
Sweet, 1950; Kippenhahn, 1958).
While turbulent convection zones provide very good prospects 
for dynamo action, one has to search for dynamo solutions in laminar
flows, if an MHD dynamo is considered an option for the presence of
magnetic fields on CP stars. Solutions for special flow constructions 
are known for more than 50~years (e.g.\ Bullard \& Gellman, 1954) and 
always provide non-axisymmetric solutions according to the theorem of 
Cowling (1934). A comprehensive compilation of spherical flows and their 
dynamo solutions was given by Dudley \& James (1989, hereafter DJ). The 
Cowling theorem says that an axisymmetric magnetic field cannot be sustained
by dynamo action. This is essential for laminar dynamos, while it is
unimportant for turbulent dynamos where nonaxisymmetries will always
be present in small-scale fields.

The rotation rate is typically normalized in terms of the magnetic Reynolds 
number ${\rm Rm}=R^2\Omega/\eta$ where $R$ is the radius of the sphere and 
$\Omega$ is a typical angular velocity. DJ constructed a very 
simple flow which does provide a dynamo for fairly low Rm. All these dynamo 
models are kinematic; the equation of motion with the corresponding Lorentz 
force is not considered in any of them.
Only the induction equation
\begin{equation}
 \frac{\partial \vec B}{\partial t} = \nabla\times\left(\vec u\times \vec B
 - \eta\nabla\times\vec B\right),
 \label{induction}
\end{equation}
with a prescribed $\vec u$. This is suitable for searches for the onset of dynamo
action; the full nonlinear evolution of the dynamo also requires the
momentum equation with a Lorentz force. All the dynamo and stability results 
obtained for this paper were computed with the MHD code by Hollerbach 
(2000). 

\begin{figure}[t]
\centerline{\includegraphics[width=9cm,clip=]{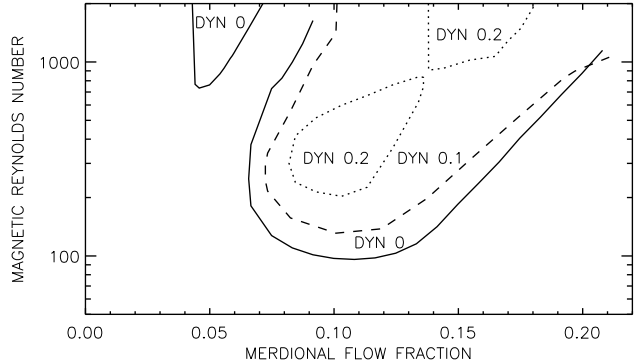}}
\caption{Dynamo action for the modified dynamo by Dudley \&
James (1989). The solid line refers to a full-sphere dynamo
where the meridional flow covers the entire sphere, the
dashed line is for a meridional going down to $0.1R_\ast$, and
the dotted line is for a flow down to $0.2R_\ast$. The regions
denoted with `DYN 0' etc.\ show dynamo action for these three
cases, respectively.}
\label{dudley_james}
\end{figure}

The original DJ flow has an $\Omega(r)$ profile decreasing with radius 
and a meridional circulation which points towards the equator near the 
surface. Figure~\ref{dudley_james} shows the critical Rm for the onset of 
dynamo action as a function of the meridional flow speed. The amplitude of 
the meridional flow velocity is measured as a fraction of the maximum azimuthal 
velocity, $\epsilon={\rm max}(u_\theta)/{\rm max}(u_\phi)$. The original 
DJ system for a full sphere is shown as a solid line. In an A star, we expect 
a convective core with very strong turbulent diffusion acting like a vacuum 
hole. We have modified the flow to obey a central hole and have plotted
the critical Rm lines for core sizes of $0.1R$ and $0.2R$. The area in 
which dynamo action is found diminishes and is entirely gone for an inner
hole of $0.24R$. Moreover, a meridional flow consistent with a negative 
$\Omega$ gradient would actually be poleward below the surface. There is 
no dynamo action for such a flow, but it is retained adding latitudinal 
differential rotation (Arlt, 2006). For another class
of meridional flows without any differential rotation, Moss (2006)
did find dynamo solutions, all with critical Rm beyond $10^4$. Since
all the solutions found so far consist of rather simple $m=1$ modes, 
the diversity of magnetic field configurations observed can probably
not be covered by such dynamos.

\section{The stability of magnetic fields in radiative envelopes}
There are essentially two types of instabilities which are relevant
for magnetic fields in radiative envelopes. One is the magnetorotational
instability (MRI; Velikhov, 1959; Balbus \& Hawley, 1991) which relies on
a differential rotation decreasing with distance from the axis,
and the other class is curent-driven instabilities for which 
rotation is only a modifier.

Especially in the beginning of the existence of the radiative envelope,
the magnetorotational instability may be of interest for mCP sars. Only 
small magnetic fields are required for the onset of the instability. 
Since the energy source for the instability is the differential rotation
instead of the magnetic field, perturbations may grow up to considerable 
strength in fairly short times. The stability analysis for nearly all 
magnetic-field configurations shows instability for field strengths which 
are not too strong (Balbus \& Hawley, 1998).
How strong is that? The instability becomes significantly weaker if 
the scale of the dominant unstable mode is larger than the object size. 
In a CP star with a differential rotation of as small as 5\%, the maximum 
field strength is as large as 20~kG; it is larger than 100~kG for 30\%
differential rotataion. Even though the MRI is a weak-field instability 
it has a huge range of applicability in a stellar context.

Arlt et al. (2003) have studied such a scenario in a spherical
shell. The setup contained a magnetic field perturbation with
both axisymmetric and non-axisymmetric parts, and without symmetry
about the equator, in order to allow for a variety of modes to grow.
The rotation rate decreases with axis distance.
The model is simplified in that it employs an incompressible
spherical shell with constant density, but allows for a subadiabatic
temperature profile as is present in radiative zones (Boussinesq
approximation). The equations now read
\begin{eqnarray}
\frac{\partial\vec{u}}{\partial t}&=& 
   \nu\nabla^2\vec{u}
  -\left(\vec{u}\cdot\nabla\right)\vec{u}
  -\nabla p
  -\alpha\Theta \vec{g}
  +\frac{1}{\mu_0\rho_0}(\nabla\times \vec{B})\times \vec{B},\label{ns}\\
\frac{\partial \vec{B}}{\partial t} &=& 
   \eta\nabla^2 \vec{B} 
  +\nabla\times(\vec{u}\times \vec{B}),\label{induction_mri}\\
\frac{\partial \Theta}{\partial t}&=&
   \kappa\nabla^2 \Theta
  -\vec{u}\cdot\nabla\Theta
  -\vec{u}\cdot\nabla T_{\rm s},\label{temperature} 
\end{eqnarray}
where $\nu$ is the kinematic viscosity, $p$ is the pressure, 
$\alpha$ is the thermal expansion coefficient, $\Theta$ denotes 
temperature deviations from a purely conductive background profile $T_{\rm s}$, 
$\vec{g}$ is the gravitational acceleration, and $\kappa$ is the
thermal diffusivity. The constants $\mu_0$ and $\rho_0$ are the
magnetic permeability and the density, respectively.


The full three-dimensional and nonlinear evolution of an 
initial rotation profile and magnetic field is followed. The
MRI sets in much faster than the diffusive time-scale, and the
fluctuations of $\vec{u}$ and $\vec{B}$ start to redistribute
angular momentum within the radiative zone by Reynolds stresses
(correlation of velocity fluctuations) and Maxwell stresses
(correlation of magnetic-field fluctuations).

\begin{figure}[t]
\centerline{\includegraphics[width=9cm,clip=]{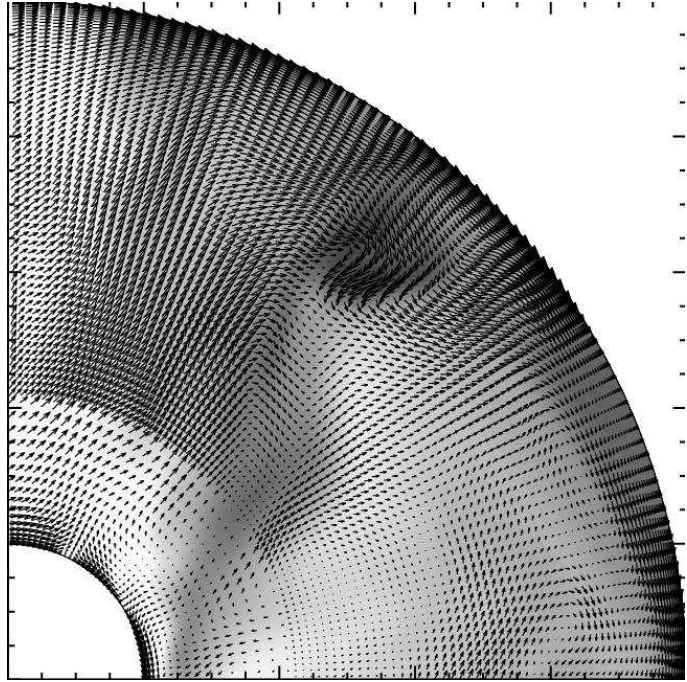}}
\caption{Velocity field after the onset of the magnetorotational instability
in a stratified sphere. The graph is a vertical cut-out of the full-sphere 
simulation.}
\label{sps_apdec_d2}
\end{figure}

The model was modified to include a background density stratification,
sacrificing the temperature equation though. The time-scale for
the redistribution of angular momentum remains the same as for the 
model with a stable temperture gradient. A snap-shot of a meridional section of the 
resulting flows is shown in Figure~\ref{sps_apdec_d2}. At that
point the angular momentum is already entirely redistributed towards
a uniform rotation. Realistic magnetic Reynolds numbers are not
accessible in a time-dependent numerical simulation. One always
has to compute a set of scenarios and extrapolate the results
such as magnetic field strength and decay time for the differential
rotation to stellar parameters. The calculations presented here
lead to a time-scale between 10 and 100~Myr for the decay of any
initial differential rotation in mCP stars.

The other class of instabilities is not based on the differential 
rotation, but on the presence of currents in the initial magnetic 
field. Nearly all poloidal magnetic fields become unstable when 
imposing a small non-axisymmetric perturbation (cf. Braithwaite, 
2007, and references therein). The same holds for purely toroidal 
magnetic fields which were studied by Vandakurov (1972) and Tayler 
(1973). 

\begin{figure}[t]
\centerline{\includegraphics[width=7cm,clip=]{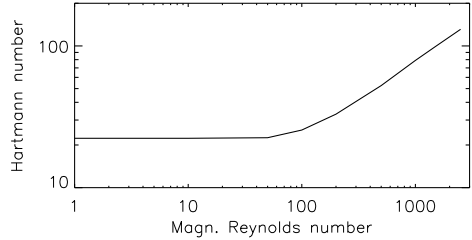}}
\caption{Stability of a poloidal magnetic field against a non-axisymmetric
perturbation in dependence on the rotation rate in terms of Reynolds number.}
\label{poloidal_field}
\end{figure}

Solid-body rotation has a stabilizing effect on poloidal as well as
toroidal magnetic fields. For this review, the dependence of the stability 
of a purely poloidal field on the rotation rate was computed. The 
normalized rotation rate is again expressed by the magnetic Reynolds 
number. The stability limit versus Rm is plotted in Fig.~\ref{poloidal_field}.
The Hartmann number is the dimensionless magnetic field strength; it
needs to be converted to physical values for specific objects.
At high Rm -- the ones in which we are interested in the case of CP stars -- 
the critical magnetic field strength becomes a power function of Rm. 
We can try to extrapolate the graph to the very high stellar magnetic 
Reynolds numbers of CP stars. The result scales to a critical field 
strength of about 10~G for a rotation period of 1~day, and to about 
1~G for a 100-day period. Maximum field strengths of up to 100~G were found 
for purely toroidal fields in the solar radiation zone (Arlt et al.,
2007). All these limits are much below the observed field strengths.

It is possible construct a magnetic field of unlimited stability.
Chandrasekhar \& Kendall (1957) reported about an entirely force-free
field, and we confirmed its stability numerically. The construction
requires the sphere to be embedded in a perfectly conducting medium.
If vacuum is employed, the stability is lost, and surface forces
emerge with unkown effect. Since the nonlinear simulations of
Braithwaite \& Nordlund (2006) lead to a configuration very similar
to the force-free field, further studies are worthwhile in this
direction.


\section{An early-life dynamo}
The early contraction phase of star formation may host a turbulent
dynamo which can provide kG-fields. Once the star becomes radiative,
the generated field could become frozen in and visible at the surface
when the convective shell has vanished. The idea is supported
by the findings of Dikpati et al. (2006) who showed that the
oscillatory dynamo of the solar convection zone can lead to 
the build-up of stationary fields in the radiative solar interior.
The following numerical experiment probes the applicability of such
a mechanism for mCP stars. A similar approach was followed by Kitchatinov
et al. (2001) for solar-type stars.

We start with a fully convective sphere in which the $\alpha$-effect
is believed to operate as a field generator. The $\alpha$-effect is
the description of the generation of poloidal magnetic fields from
toroidal ones and vice versa in rotating, stratified turbulence.
As time progresses, the convection zone becomes a shell, and the
inner radius of the shell increases gradually. The core is radiative
which is represented numerically by a vanishing $\alpha$-effect and
a magnetic diffusivity which is 1000~times smaller than the turbulent
magnetic diffusivity $\eta_{\rm T}$ in the convection zone. The 
induction equation for a dynamo with a magnetic diffusivity depending
on the location in the star is
\begin{equation}
 \frac{\partial \vec B}{\partial t} = \nabla\times\left(\vec u\times \vec B
 +\alpha \vec B - \eta\nabla\vec B - \frac{1}{2}(\nabla\eta)\times \vec B\right),
 \label{induction_alpha}
\end{equation}
where $\vec u$ can contain all large-scale velocities such as differential 
rotation and meridional circulation. The $\alpha$ effect is distributed 
such that it is proportional to $\sin^2\theta\cos\theta$ in the convective 
shell as well as in the inner convective core forming near the end of the 
simulation, where $\theta$ is the colatitude. 
We also introduced an anisotropy in the $\alpha$-tensor as 
expected from rotating turbulence, such that the $\alpha$-component 
parallel to the rotation axis vanishes (R\"udiger \& Kitchatinov, 1993).

The last term in Eq.~(\ref{induction_alpha}) is due to the turbulence
gradient and can be described as turbulent pumping since it has the 
effect of a velocity (Zel'dovich, 1957; Ruzmaikin \& Vainshtein, 1978). 
The term is often neglected in computations of turbulent dynamos with 
functions $\eta_{\rm T}(r)$, but plays a significant role in confining 
magnetic fields in radiative interiors.

\begin{figure}[t]
\centerline{\includegraphics[width=11cm,clip=]{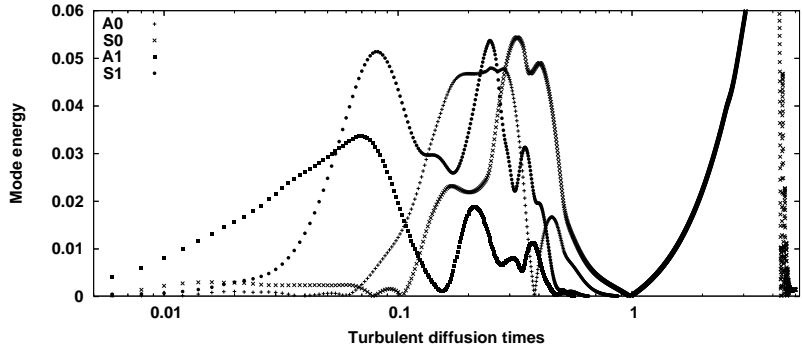}}
\caption{Dynamo solution based on a an evolution from a fully convective
sphere to a radiative sphere with convective core. `A' refers to
equator-antisymmetry, `S' to symmetry. `0` denotes 
axisymmetric $m=0$ modes, `1' non-axisymmetric $m=1$ modes.}
\label{alpha2omega_evol_ap}
\end{figure}

The temporal evolution in Fig.~\ref{alpha2omega_evol_ap} shows a short 
phase with dominance of the non-axisymmetric mode. This agrees with 
earlier findings that fully convective shells may provide $m=1$ dynamo 
modes. When the radiative core becomes larger than $0.22R_\odot$, the 
$m=0$ dominates and does so for the remaining time of the simulation. 
The storage of magnetic flux in the radiative interior appears to
be chief\/ly axisymmetric, unfortunately. An initial convective dynamo 
may thus not provide the observed fields directly. The onset of a core
dynamo does produce strong internal fields, but these weaken severely
towards larger distances from the center. This was also found from direct
numerical simulations by Brun et al. (2005).

\section{Fossil fields}
An issue we have not addressed in this review so far is the survival of large 
scale magnetic fields from before the collapse of the star. Although we 
concentrated on dynamos and instabilities, a few words may be necessary for 
a comprehensive picture. A fully convective phase suggested to take place 
for stars with masses below $2.4~M_\odot$ (Palla \& Stahler, 1993) could 
destroy these primordial fields in a rather short time of a few thousand 
years due to turbulent diffusion. Moss (2003) argued, however, that the
exceptional strength of the fields in mCP star progenitors reduces
turbulent diffusion and allows the survival of sufficient flux to be
obervable after the star has reached the zero-age main-sequence. The
exact behaviour of the diffusion reduction -- known as $\eta$-quenching
in mean-field electrodynamics -- is not known though.

Even if it is not largely convective, the proto-stellar object will also 
be subject to differential rotation. Torques exerted by accreted material, 
magnetic coupling to the accretion disk and magnetic winds very likely cause 
differential rotation in the star (St\c{e}pie\'{n}, 2000). Primordial fields of 
arbitrary direction will excite the magnetorotational instability as long as the 
magnetic fields are weaker than $O(100)$~kG which is probably true. The
instability is actually a good candicate for the diversity if field
configurations afterwards; it also allows for amplification of the field,
with the energy being taken from the shear.


\section{Summary}
A compilation of a few possible mechanism in CP star magnetism with
their pros and contras are listed in Table~\ref{summary}. When searching 
for mechanisms explaining considerable magnetic fields in radiative 
envelopes, one always returns to the question why only a small fraction 
of A stars shows such fields. The effect of different star formation
environments on the existence of strong fields on main-sequence stars
now needs to be investigated. Bridging collapse and pre-main-sequence
evolution is not a trivial task though.

\begin{table}
\caption{Four scenarios of CP star magnetism.\label{summary}}
\begin{tabular}{|l|l|}
\hline
{\bf Background field amplified at}         & {\bf Dynamo in radiative zone}\\
{\bf star formation}                        & $+$ straight-forward mechanism \\
$+$ random orientation                      & $-$ $m=1$ modes cause regular obliquity \\
$+$ field strength decreases with age       & $-$ field strength does not decreases\\
$-$ convective phase or MRI produce         & \hspace{4mm}with age\\
\hspace{4mm}small-scale field               & $-$ rather independent of star formation\\
$-$ contradicts possible late emer-         & \hspace{4mm}$\rightarrow$ no distinction between A and Ap\\
\hspace{4mm}gence of field                  & $-$ dynamo difficult to excite\\
\hline
{\bf Early-life dynamo remnants}            & {\bf Early-life dynamo + field instab.}\\
$+$ field strength decreases with age       & $+$ random orientations at emergence\\
$-$ fully convective phase questionable     & $+$ field strength decreases slowly \\
$-$ non-axisymmetry only very shortly       & \hspace{4mm}with age\\
\hline
\end{tabular}
\end{table}

\end{document}